\documentclass[12pt]{iopart}
\usepackage{iopams}
\usepackage{graphicx}
\usepackage[T1]{fontenc}
\usepackage{ae}
\usepackage[latin1]{inputenc}
\usepackage{color}

\newtheorem{prop}{Proposition}

\begin{document}

%\begin{center}\today\end{center}

\title[Magnetic work?]
{The magnetic field does not perform work - or does it?}

\author{Heinz-J\"urgen Schmidt$^1$
 \footnote[3]{Correspondence should be addressed to
hschmidt@uos.de} and Thomas Br\"ocker$^1$}

\address{$^1$ Universit\"at Osnabr\"uck, Fachbereich Mathematik/Informatik/Physik,
Barbarastr. 7, D - 49069 Osnabr\"uck, Germany }

\begin{abstract}
We investigate the question discussed in the literature as to whether the magnetic field can perform work using two models that describe interacting magnetic dipoles. In the first model, the dipoles are realized by rigidly rotating charge clouds, whereas in the second model, one of the two dipoles is described by a real macroscopic spin density. The theoretical foundations of the second model are formulated in a recently published paper. We obtain equations of motion and detailed energy balance equations for both cases, but the answer to the title question depends on the choice of  criteria for ``magnetic work".

\end{abstract}

\maketitle

%%%%%%%%%%%%%%%%%%%%%%%%%%%%%%%%%%%%%%%%%%%%%%%%%%%%%%%%%%%%%%%%%%%%%%%%%%%%%%%%%%%%%%%%%%%%%%%%%%%%%%%%%%%%%%%%%%%%%%%%%%%%%%%%%%%%%%%%%%%%%%%%%%
\section{Introduction}
\label{sec:I}
%%%%%%%%%%%%%%%%%%%%%%%%%%%%%%%%%%%%%%%%%%%%%%%%%%%%%%%%%%%%%%%%%%%%%%%%%%%%%%%%%%%%%%%%%%%%%%%%%%%%%%%%%%%%%%%%%%%%%%%%%%%%%%%%%%%%%%%%%%%%%%%%%%%%%

According to a widespread opinion it holds that
``the magnetic field does no work, since the
magnetic force is perpendicular to the velocity,"
see, e.~g., \cite{J90}, section 6.7.
On the other hand, you can accelerate a magnet through the field of another magnet,
and do work in the process. Isn't that a contradiction?

This problem has been discussed frequently in the literature, see, e.~g.,  \cite{M74,C79,OA13,G14,VSR23} and on internet platforms.
Recently it has been taken up again by J.~A.~Barandes \cite{B19,B21},
who considers an electromagnetic field coupled to a relativistic charged particle with spin and derives expressions for the force on this particle which include the magnetic field. Hence he answers the question "Can the magnetic field do work?" in the affirmative. Our work can be seen as a complementary contribution to the discussion in \cite{B19,B21}, as we will elaborate below.

The mentioned paradox refers to the energy exchange between the electromagnetic field and matter. If we want to analyze it, it is inevitable to include theoretical descriptions of matter besides Maxwell's equations for the electromagnetic field, even if one restricts oneself to strong simplifications. Electrodynamics knows only the reduced description of matter in terms of charges and currents (plus possibly spin densities, see below), but has no possibility to represent interactions and energy flows within matter. For a theoretical description of matter, in principle, both classical or relativistic mechanics and quantum mechanics come into question.

But first we want to reduce the paradox to its essential core.
The situation where two magnets attract (or repel) each other and
perform work in the process shall be described simplistically
by the interaction of two magnetic dipoles, a ``small"  and a ``large" one.
Suitable assumptions are made to ensure that the large dipole retains its position and magnetic moment,
so that its magnetic field remains constant over time and can be considered as the ``external" field.
The restriction to temporally constant external fields makes sense,
as time-variable magnetic fields are associated with electric fields,
which generally lead to a flow of energy between the field and matter.
In this sense, time-dependent magnetic fields perform work,
which is completely undisputed in the above-mentioned debate.
The question there is rather whether time-constant magnetic fields perform work.

As a further simplification, let us refrain from considering fully relativistic equations including radiation effects caused by accelerated charges.
The force effects contributing to the apparent paradox can already be calculated non-relativistically and quasi-statically to a good approximation.
Correspondingly, we will neglect relativistic effects in the motion of the small dipole and adopt a description in terms of classical mechanics.

\begin{figure}[ht!]
\centering
\includegraphics*[clip,width=0.7\columnwidth]{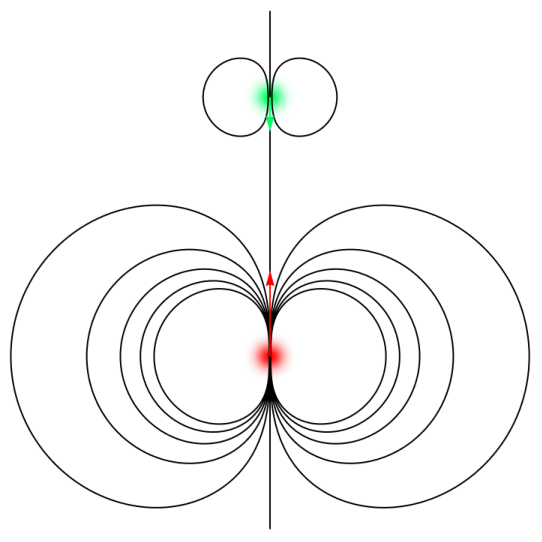}
\caption{Sketch of two magnetic dipoles realized by rotating clouds, a ``large" one (red color) and a ``small" one (green color), with opposite magnetic
moments oriented parallel to the vertical $z$-axis.
}
\label{FIGSL}\end{figure}

There are two possibilities for the theoretical description of the two magnetic dipoles, which now stand for ``matter". In the approach commonly used in electrodynamics, the magnetic moments are realized by current distributions (model A, or  ``Amp\`{e}re model"),  see Section \ref{sec:A},  The well-known equation for the force on a magnetic dipole in a weakly inhomogeneous magnetic field can be reproduced if we assume the current distribution of the small dipole to be sharply localized,
see subsection \ref{sec:AF}.
In order to obtain the simplest possible description in terms of classical mechanics, we will realize the dipoles by rigidly rotating charged clouds with, say, negative charge and a rotationally invariant Gaussian density. Both magnetic moments are initially, and hence for all times, oriented parallel to the $z$-axis, see Figure \ref{FIGSL}.
 Since we do not want to consider any disturbing effects caused by the electric fields of the charged clouds, we additionally assume rigid clouds with the opposite, positive charge density, which have the same center of mass as the negatively charged clouds and initially do not rotate. The center of mass of the two clouds representing the small dipole will move into $z$-direction under the influence of the force exerted by the magnetic field of the large dipole.
 see subsection \ref{sec:AG}.
 In summary, we therefore consider four clouds that realize two electrically neutral dipoles in pairs.

%===================    figure   =================================
%===================    figure  =======

However, this description is unrealistic when dealing with magnetic dipoles that are physically caused in whole or in part by spin densities. Therefore, we will treat this case separately, see section \ref{sec:B}, relying on a recently published proposal \cite{SB23} to supplement the classical Maxwell equations with spin densities as possible sources of magnetic fields (model B). However, it would not be appropriate to describe also the large dipole by a real spin density. In general, both magnetic moments would precess around each other and the magnetic field of the large dipole could not be kept constant in time.
This can be illustrated by studying the interaction of two magnetic dipoles fixed in space, see \cite{SSHL15}.
Therefore, we will describe the large dipole by two clouds, one of which is rotating, as in model A, ensuring that the position
and angular velocities of these clouds remain constant in time, see subsection \ref{sec:BG}.

Only the small dipole is represented by a rigid, electrically neutral cloud with a genuine magnetic moment density. Its center of mass can move in the $z$-direction, but it turns out that the cloud will not rotate. Also in model B the small dipole will be extended and not considered as a point dipole, which has the advantage that its field energy will be finite.

Therefore, our treatment of the problem is complementary to the discussion in \cite{B21}, since we (1) consider only a case study and not a complete theory,
(2) restrict ourselves to non-relativistical calcuations and (3) consider extended dipoles with finite field energy.

Interestingly, the analysis of the paradox is different in the two cases mentioned. In both cases, we derive equations of motion (eom) for the position of the small dipole, and, in case A, for the angular velocities of its two clouds,
see subsection \ref{sec:AE} and \ref{sec:A2}.
For these eom and for the energy balance, it is necessary to take into account the electric field $\mathbf{E}$ induced by the temporal change of the magnetic field of the small dipole.
If the two magnetic dipoles are described by rigidly rotating charged clouds, the argument cited at the beginning comes into full effect, namely that the Lorentz force is perpendicular to the velocity of each charge element and therefore cannot perform any work. However, this does not apply to the $\mathbf{E}$ field, which causes an exchange of energy between the two dipoles and the field. It can be proven that the sum of the field energy and the total kinetic energy of the mechanical system remains constant,
see subsection \ref{sec:AB}.
Surprisingly, the kinetic rotational energy of the large dipole
is variable and its change  must also be taken into account.
Nevertheless, if the mass (and thus the moment of inertia) of the large dipole is chosen
 to be very large compared to the mass of the small dipole,
 this result is compatible with the assumption of a constant external field.

If, on the other hand, we are dealing with ``real" magnetic dipoles, a mechanical description using a classical spin density and a so-called Zeeman term, which acts like a potential energy, is appropriate,
see the subsections \ref{sec:BG} and \ref{sec:BF}.
The Zeeman term is not ad hoc, but can be justified, for example, by the minimal coupling of a Dirac spinor to a classical electromagnetic field, see \cite{BD64}. In this case, case B, there is an energy exchange between kinetic translational energy and Zeeman energy without tapping the field energy. On the other hand, the field energy is not conserved but increases at the expense of the kinetic rotational energy of the large dipole, similar to model A,
see the subsection \ref{sec:BB} and \ref{sec:A1}.
We emphasize that this result is restricted to the case of (anti-)parallel magnetic moments of the two dipoles and may be more complicated in the general case where the moment of the small dipole precesses around the external magnetic field.

This results in a, perhaps unexpected, answer to the title question ``The magnetic field does not perform work - or does it?": It depends, the question must be made more precise. In case A, there is an exchange of magnetic field energy and kinetic energy of both dipoles. This could be described as ``magnetic work"; however, the energy exchange is mediated by the electric field.
In case B, the acceleration of the small dipole is associated with a decrease in Zeeman energy,
which in turn depends on the external magnetic field.
In addition, magnetic field energy and rotational energy of the large dipole are converted into each other.
But here, too, the total magnetic field is time-dependent and the energy flow is mediated by the induced electric field.
In any case, we can state that the energetic relations and the motion of the small dipole strongly depend on whether we use model A or B.
This will be outlined in the Summary.

%%%%%%%%%%%%%%%%%%%%%%%%%%%%%%%%%%%%%%%%%%%%%%%%%%%%%%%%%%%%%%%%%%%%%%%%%%%%%%%%%%%%%%%%%%%%%%%%%%%%%%%%%%%%%%%%%%%%%%%%%%%%%%%%%%%%%%%%%%%%%%%%%%
\section{Model A}
\label{sec:A}
%%%%%%%%%%%%%%%%%%%%%%%%%%%%%%%%%%%%%%%%%%%%%%%%%%%%%%%%%%%%%%%%%%%%%%%%%%%%%%%%%%%%%%%%%%%%%%%%%%%%%%%%%%%%%%%%%%%%%%%%%%%%%%%%%%%%%%%%%%%%%%%%%%

%%%%%%%%%%%%%%%%%%%%%%%%%%%%%%%%%%%%%%%%%%%%%%%%%%%%%%%%%%%%%%%%%%%%%%%%%%%%%%%%%%%%%%%%%%%%%%%%%%%%%%%%%%%%%%%%%%%%%%%%%%%%%%%%%%%%%%%%%%%%%%%%%%
\subsection{General notations and assumptions}
\label{sec:AG}
%%%%%%%%%%%%%%%%%%%%%%%%%%%%%%%%%%%%%%%%%%%%%%%%%%%%%%%%%%%%%%%%%%%%%%%%%%%%%%%%%%%%%%%%%%%%%%%%%%%%%%%%%%%%%%%%%%%%%%%%%%%%%%%%%%%%%%%%%%%%%%%%%%

As explained in the Introduction, we model each of the two dipoles by two rigidly rotating clouds with opposite charge densities.
These four clouds are distinguished by a superscript $(0)$ for the small dipole or $(1)$ for the large dipole and additionally,
by a subscript $\pm$, which refers to the total positive/negative charge of the cloud.
The center of mass of either dipole will be denoted by ${\mathbf r}_0^{(i)},\,i=0,1,$ and can be chosen of the special form
\begin{equation}\label{specialr}
 {\mathbf r}_0^{(1)}=\mathbf{0} \quad\quad\mbox{and}\quad {\mathbf r}_0^{(0)}=(0,0,z_0(t))\quad\mbox{such that}\quad z_0(0)=a>0
 \;,
\end{equation}
anticipating the result that the motion of the small dipole is restricted to the $z$-axis.
The various densities are proportional to a normalized, rotationally invariant, Gaussian density
\begin{equation}\label{Gaussian}
 \rho(r)=\frac{\alpha^3}{\pi^{3/2}}{\sf e}^{-\alpha^2\,r^2}
 \;,
\end{equation}
depending on a parameter $\alpha>0$. For the radius $r$ we have to insert
\begin{equation}\label{rinsert}
r=\left|{\mathbf r} -{\mathbf r}_0^{(i)}(t)\right|
\end{equation}
depending on which cloud is considered.
If necessary, we distinguish between the two dipoles by writing $\rho^{(0)}$ or $\rho^{(1)}$.
For later use we note
\begin{equation}\label{gradrho}
 \nabla\,\rho = -2\alpha^2 {\mathbf r}\,\rho
 \;.
\end{equation}

The four clouds are rotating with angular velocity $\omega_{\pm}^{(i)}(t)$ around the $z$-axis,
where for the large dipole we set
\begin{equation}\label{specialomega}
  \omega_{+}^{(1)}(t)=0  \quad\quad\mbox{and}\quad  \omega_{-}^{(1)}(t)=\mbox{const.}
\end{equation}
Moreover, the clouds will have mass densities
\begin{equation}\label{massdensity}
 M_\pm^{(i)}\,\rho\left(\left|{\mathbf r} -{\mathbf r}_0^{(i)}(t)\right|\right)
 \;,
\end{equation}
charge densities
\begin{equation}\label{chargedensity}
 e_\pm^{(i)}\,\rho\left(\left|{\mathbf r} -{\mathbf r}_0^{(i)}(t)\right|\right)\quad\quad\mbox{satisfying}\quad e_-^{(i)}=-e_+^{(i)}<0
 \;,
\end{equation}
and magnetic moment densities
\begin{eqnarray}\label{magmomdensity1}
 {\boldsymbol\mu}_\pm^{(i)}&=&\left(\begin{array}{c}
  0 \\
  0 \\
  m_\pm^{(i)} (t)\\
\end{array}
\right)
\,\rho\left(\left|{\mathbf r} -{\mathbf r}_0^{(i)}(t)\right|\right)\\
\label{magmomdensity2}
&=&
\frac{e_\pm^{(i)}}{2\alpha^2}
\left(\begin{array}{c}
  0 \\
  0 \\
  \omega_\pm^{(i)} (t)\\
\end{array}
\right)
\,\rho\left(\left|{\mathbf r} -{\mathbf r}_0^{(i)}(t)\right|\right)
\;.
\end{eqnarray}
The total magnetic dipole moments of the respective dipoles are designated as:
\begin{equation}\label{totalmagmom}
 {\mathbf m}^{(i)}=  {\mathbf m}_+^{(i)}+ {\mathbf m}_-^{(i)}=\left(\begin{array}{c}
  0 \\
  0 \\
 m^{(i)} (t)\\
\end{array}
\right),\quad\mbox{for } i=0,1
 \;.
\end{equation}
Recall that $m^{(1)}(t)=\mbox{const.}$.
To simplify the discussion we choose
the case where the two dipoles repel, say,
\begin{equation}\label{m1}
m^{(1)}>0\quad\quad \mbox{and}\quad m^{(0)}<0
\;.
\end{equation}

%%%%%%%%%%%%%%%%%%%%%%%%%%%%%%%%%%%%%%%%%%%%%%%%%%%%%%%%%%%%%%%%%%%%%%%%%%%%%%%%%%%%%%%%%%%%%%%%%%%%%%%%%%%%%%%%%%%%%%%%%%%%%%%%%%%%%%%%%%%%%%%%%%
\subsection{Force and torque}
\label{sec:AF}
%%%%%%%%%%%%%%%%%%%%%%%%%%%%%%%%%%%%%%%%%%%%%%%%%%%%%%%%%%%%%%%%%%%%%%%%%%%%%%%%%%%%%%%%%%%%%%%%%%%%%%%%%%%%%%%%%%%%%%%%%%%%%%%%%%%%%%%%%%%%%%%%%%

The magnetic field generated by a cloud with charge density $e\,\rho(r)$, its center of mass located at ${\mathbf r}_0=\mathbf{0}$,
and rotating with angular velocity ${\boldsymbol\omega}=(0,0,\omega(t))$ is given by
\begin{eqnarray}\label{magneticfield}
{\mathbf B}&=&
\frac{e \mu _0 \omega (t)}{4 \pi ^{3/2} \alpha  r^4}\,{\sf e}^{-\alpha^2 r^2}\, {\mathbf B}^{(a)}+
\frac{e \mu _0 \omega (t)}{8 \pi  \alpha ^2 r^5}\,\mbox{erf}(\alpha r)\, {\mathbf B}^{(b)}\;,\\
\nonumber
\mbox{where}&&\\
\label{magneticfielda}
 {\mathbf B}^{(a)}&=&
 \left(
\begin{array}{c}
 -x z \left(2 \alpha ^2 r^2+3\right) \\
 -y z \left(2 \alpha ^2 r^2+3\right) \\
 2 r^2 \left(\alpha ^2 \left(x^2+y^2\right)-1\right)+3 \left(x^2+y^2\right) \\
\end{array}
\right)
\;,
 \\
 \label{magneticfieldb}
 {\mathbf B}^{(b)}&=&
 \left(
\begin{array}{c}
 3 x z \\
 3 y z \\
 3 z^2 -r^2  \\
\end{array}
\right)
\;.
\end{eqnarray}
We have derived (\ref{magneticfield}--\ref{magneticfieldb})
by different methods, one being described in \cite{M82}, and have confirmed it by checking the
equations $\nabla\cdot {\mathbf B}=0$ and $\nabla\times {\mathbf B}=\mu_0\,{\mathbf j}=\mu_0 e \rho\,{\boldsymbol\omega}\times{\mathbf r}$.
This result will be used to calculate the magnetic field ${\mathbf B}^{(1)}:={\mathbf B}_-^{(1)}$ generated by the large dipole
and ${\mathbf B}^{(0)}:={\mathbf B}_+^{(0)}+{\mathbf B}_-^{(0)}$ generated by the two clouds of the small dipole.

For large distances from the center of mass of the rotating cloud, i.~e., for $\alpha\,r\to\infty$
we have $ {\sf e}^{-\alpha ^2 r^2}\to 0$ and $\mbox{erf}(\alpha  r)\to 1$.
Then (\ref{magneticfield}) asymptotically assumes the form of the field of a magnetic point dipole.
This asymptotic form will always be used when calculating the force or the torque exerted from the large dipole upon the small one.
Moreover, for these calculations we will use the linear approximation of ${\mathbf B}^{(1)}$ at the position ${\mathbf r}_0^{(0)}(t)$ of the small dipole.

The result of the action of the external field ${\mathbf B}^{(1)}$ on the small dipole is, firstly, a total force given,
in accordance with the textbook equation, by
\begin{equation}\label{force}
 {\mathbf F}^{(0)}= \nabla\left( {\mathbf m}^{(0)}\cdot{\mathbf B}^{(1)}\right)=\left(0,0,-\frac{3\, B\,m^{(0)}}{a+z_0(t)}\right)
 \;,
\end{equation}
where $B$ is the $z$-component of ${\mathbf B}^{(1)}$ in the point dipole approximation at the position of the small dipole, i.~e.,
\begin{equation}\label{B}
B=\frac{\mu _0 m^{(1)}}{2 \pi  (a+z_0(t))^3}>0
\;.
\end{equation}
In this context it is remarkable that T.~H.~Boyer derives two different expressions for the force (\ref{force}),
depending on whether one chooses the electric current model or the separated magnetic charge model \cite{B88}, see also \cite{V90}.
However, if $\nabla\times{\mathbf B}^{(1)}={\mathbf 0}$, which is approximately satisfied in our case, both expressions coincide.

\begin{figure}[ht!]
\centering
\includegraphics*[clip,width=1.0\columnwidth]{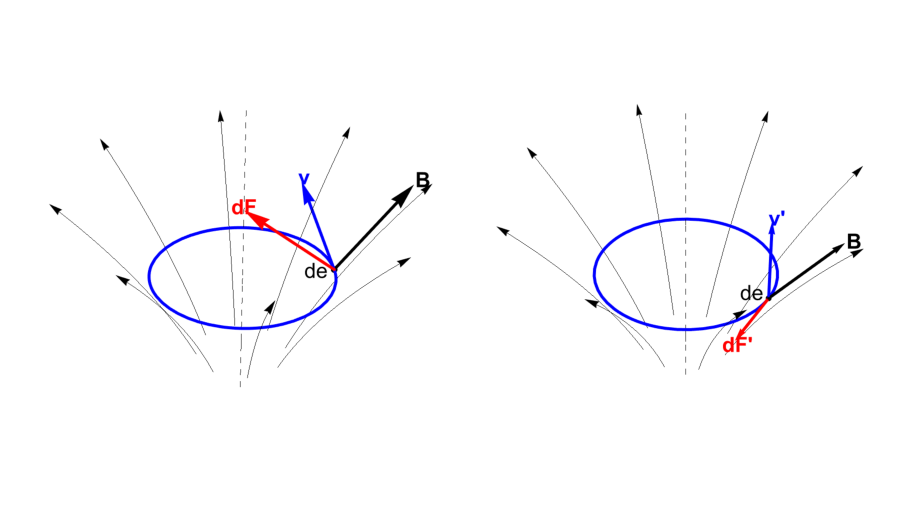}
\caption{Left panel: Force $d{\mathbf F}= de\,{\mathbf v}\times {\mathbf B}$
on a charge element $de<0$ rotating anti-clockwise on the blue circle about the z-axis (dashed black line).
The initial velocity ${\mathbf v}$ is tangent to the blue circle and the force $d{\mathbf F}$ has a constant z-component.\\
Right panel: Force $d{\mathbf F}'= de\,{\mathbf v}'\times {\mathbf B}$
on a charge element $de<0$, with the constant component ${\mathbf v}'$ of the velocity into z-direction.
Consequently, $d{\mathbf F}'$ represents the  constant tangential component of $d{\mathbf F}$.
}
\label{FIGR12}\end{figure}

Secondly, the torque ${\mathbf N}^{(10)}$ exerted by  ${\mathbf B}^{(1)}$ on the small dipole is
calculated as
\begin{equation}\label{torque}
 {\mathbf N}^{(10)}=(0,0,\frac{3 B m^{(0)} \dot{z}_0(t)}{a+z_0(t)})
\;,
\end{equation}
This torque leads to a slowdown in the rotation of the negatively charged cloud and thus to a reduction in its magnetic moment.
For a qualitative explanation of these results see Figure \ref{FIGR12}.

%===================    figure   =================================
%===================    figure  =======

The induced electric vortex fields ${\mathbf E}_\pm^{(0)}$ are obtained as ${\mathbf E}_\pm^{(0)}=-\frac{\partial}{\partial t}{\mathbf A}_\pm^{(0)} $,
where ${\mathbf A}_\pm^{(0)}$ is the vector potential of ${\mathbf B}_\pm^{(0)}$, the latter calculated according to  (\ref{magneticfield}--\ref{magneticfieldb}).
Note that ${\mathbf E}_\pm^{(1)}={\mathbf 0}$ since ${\mathbf B}_\pm^{(1)}$ is assumed to be constant in time.
The final result reads
\begin{eqnarray}
\label{efield}
  {\mathbf E}_\pm^{(0)} &=& \omega_\pm^{(0)} (t) \left(z-z_0(t)\right) \dot{z}_0(t) {\mathbf U}+ \dot{\omega}_\pm^{(0)}(t) {\mathbf V}\;, \\
  \nonumber
  \mbox{where}&&\\
  {\mathbf U} &=& \left(
  \begin{array}{c}
   y \left(3 \sqrt{\pi } \mbox{erf}(\alpha  w)-2 \alpha  w {\sf e}^{-\alpha ^2 w^2} \left(2 \alpha ^2
   w^2+3\right)\right) \\
   - x\left(3  \sqrt{\pi } \mbox{erf}(\alpha  w)-2 \alpha  w  {\sf e}^{-\alpha ^2 w^2} \left(2 \alpha ^2
   w^2+3\right)\right) \\
   0
  \end{array}
  \right)\;,
   \\
  {\mathbf V} &=&  \left(
  \begin{array}{c}
  y\left(w^2 \left(\sqrt{\pi } \mbox{erf}(\alpha  w)-2 \alpha  w  {\sf e}^{-\alpha ^2 w^2} \right)\right)\\
  -x\left(w^2 \left(\sqrt{\pi } \mbox{erf}(\alpha  w)-2 \alpha  w  {\sf e}^{-\alpha ^2 w^2} \right)\right) \\
   0
  \end{array}
  \right)\;,\\
  \nonumber
  \mbox{and}&&\\
  w&=& \sqrt{r^2+z_0(t){}^2-2 z z_0(t)}
  \;.
\end{eqnarray}

The total force exerted by the electric fields  ${\mathbf E}_\pm^{(0)}$ on the two clouds of the small dipole vanishes,
but there will appear some non-zero torque ${\mathbf N}_\pm^{(00)}$ that leads to a change of ${\boldsymbol\omega}_\pm^{(0)}$
according to the equation ${\mathbf N}=\theta\,\dot{\boldsymbol\omega}$ and using the values of the inertial moments
\begin{equation}\label{inertialmoment}
 \theta_\pm^{(0)} =\frac{ M_\pm^{(0)}}{\alpha^2}
 \;.
\end{equation}
of the two clouds representing the small dipole.

Altogether, the above considerations result in a coupled system of
differential equations for the motion of the magnetic dipole,
where the variables are $z_0(t)$ and $\omega_\pm^{(0)}(t)$.
This system represents a genuine {\em interaction} between matter and field.
In most physical applications, only a one-sided action is considered,
either the action of the given electromagnetic fields on the charges and currents,
or the generation of electromagnetic fields by given charges and currents.
In our case, the two charged clouds move in the external magnetic field and their
movement generates an electric field which acts back on them, and so on.

%%%%%%%%%%%%%%%%%%%%%%%%%%%%%%%%%%%%%%%%%%%%%%%%%%%%%%%%%%%%%%%%%%%%%%%%%%%%%%%%%%%%%%%%%%%%%%%%%%%%%%%%%%%%%%%%%%%%%%%%%%%%%%%%%%%%%%%%%%%%%%%%%%
\subsection{Equation of motion}
\label{sec:AE}
%%%%%%%%%%%%%%%%%%%%%%%%%%%%%%%%%%%%%%%%%%%%%%%%%%%%%%%%%%%%%%%%%%%%%%%%%%%%%%%%%%%%%%%%%%%%%%%%%%%%%%%%%%%%%%%%%%%%%%%%%%%%%%%%%%%%%%%%%%%%%%%%%%

It is advisable to formulate the equation of motion (eom) in dimensionless variables.
For this purpose, we introduce the following problem-related physical units.
\begin{itemize}
  \item Length:\quad $a$ (initial distance between the large and the small dipole)
  \item Mass: \quad $M_-^{(0)}$ (the mass of the negative cloud of the small dipole)
  \item Charge:\quad $e=|e_-^{(0)}|$ (the absolute charge of the negative cloud of the small dipole)
  \item Time: \quad $1/\omega_1$ (inverse Larmor frequency).
\end{itemize}
The Larmor frequency $\omega_1$ is given by
\begin{equation}\label{Larmor}
 \omega_1=\frac{e\,B}{M_-^{(0)}}= \frac{e \mu_0 m^{(1)}}{2\pi a^3 M_-^{(0)} }
 \;.
\end{equation}
Further we can form the following dimensionless ratios:
\begin{eqnarray}
\label{ratios1}
  \epsilon &:=& \frac{1}{\alpha\,a}\;, \\
  \label{ratios2}
  \mu_1 &:=& \frac{M_-^{(0)}}{M_+^{(0)}}\;, \\
  \label{ratios3}
  \mu_2 &:=& \frac{M_-^{(0)}}{e^2\,\mu_0\,\alpha}\;, \\
  \label{ratios4}
  \gamma_1 &:=& \frac{m^{(0)}}{m^{(1)}}\;, \\
  \label{ratios5}
  \frac{\omega_0}{\omega_1} &=& \frac{4\pi \gamma_1 \mu_2}{\epsilon^3}
  \;.
\end{eqnarray}
In (\ref{ratios5}) we have introduced the initial angular velocity $\omega_0:=\omega_-^{(0)}(0)$ of the  negative cloud of the small dipole.
Note that $\gamma_1<0$ for the case of anti-parallel magnetic moments of both clouds as chosen in this paper.
The eom for the dimensionless variables (denoted by the same letters and omitting the superscript $(0)$) then read:
\begin{eqnarray}
\label{eom1}
  \ddot{z}_0(t) &=& \frac{3 \mu _1 \epsilon ^2 }{2 \left(\mu _1+1\right)}\,
  \frac{\omega _-(t)-\omega _+(t)}{ \left(z_0(t)+1\right)^4}
    \;,\\
   \label{eom2}
  \dot{\omega}_+(t) &=& \frac{36 \pi ^{3/2} \mu _1 \mu _2}{\left(\sqrt{2} \mu _1+24 \pi ^{3/2} \mu
   _2+\sqrt{2}\right)}
    \frac{ \dot{z}_0(t)}{ \left(z_0(t)+1\right)^4}
   \;, \\
   \label{eom3}
  \dot{\omega}_-(t) &=& -\frac{36 \pi ^{3/2} \mu _2}{\left(\sqrt{2} \mu _1+24 \pi ^{3/2} \mu _2+\sqrt{2}\right)}
  \frac{ \dot{z}_0(t)}{ \left(z_0(t)+1\right)^4}
   \;.
\end{eqnarray}
The equations (\ref{eom2}) and (\ref{eom3}) can be integrated, using the initial conditions
$\omega_+(0)=0$ and $\omega_-(0)=\omega_0$, and inserted into (\ref{eom1}) with the result
\begin{eqnarray}\nonumber
\ddot{z}_0(t) &=& {\textstyle\frac{3 \epsilon ^2}{2 \left(\frac{1}{\mu _1}+1\right) \left(z_0(t)+1\right)^7}}\\
 \label{eom4}
&&
{\textstyle \left(\omega _0+\frac{\left(\sqrt{2} \left(\mu _1+1\right) \omega _0-12 \pi ^{3/2} \left(\mu _1-1\right) \mu
   _2\right) z_0(t) \left(z_0(t) \left(z_0(t)+3\right)+3\right)}{\sqrt{2} \left(\mu _1+1\right)+24
   \pi ^{3/2} \mu _2}\right)
   \;.
}
\end{eqnarray}

It is not analytically solvable but can be studied numerically.

%%%%%%%%%%%%%%%%%%%%%%%%%%%%%%%%%%%%%%%%%%%%%%%%%%%%%%%%%%%%%%%%%%%%%%%%%%%%%%%%%%%%%%%%%%%%%%%%%%%%%%%%%%%%%%%%%%%%%%%%%%%%%%%%%%%%%%%%%%%%%%%%%%
\subsection{Energy balance}
\label{sec:AB}
%%%%%%%%%%%%%%%%%%%%%%%%%%%%%%%%%%%%%%%%%%%%%%%%%%%%%%%%%%%%%%%%%%%%%%%%%%%%%%%%%%%%%%%%%%%%%%%%%%%%%%%%%%%%%%%%%%%%%%%%%%%%%%%%%%%%%%%%%%%%%%%%%%

As a first approach to the energy balance we consider the Poynting theorem in its quasi-static approximation:
\begin{equation}\label{Poynting}
 \frac{d}{d t} \int \frac{1}{2\mu_0}\left| {\mathbf B}\right|^2\,dV = -\int {\mathbf j}\cdot {\mathbf E}\,dV
 \;,
\end{equation}
which can be derived in the usual way from the quasi-static Maxwell equations
\begin{eqnarray}
\label{quasistatic1}
  \nabla \times {\mathbf E}+\frac{\partial}{\partial t}{\mathbf B} &=& {\mathbf 0}\;, \\
  \label{quasistatic2}
  \nabla\cdot {\mathbf B}&=& 0\;, \\
  \label{quasistatic3}
  \nabla\cdot {\mathbf E}&=& 0\;, \\
  \label{quasistatic4}
  \nabla \times {\mathbf B} &=& \mu_0\,{\mathbf j}
  \;.
\end{eqnarray}
Here ${\mathbf B}={\mathbf B}^{(1)}+{\mathbf B}^{(0)}$ and
${\mathbf E}={\mathbf E}_+^{(0)}+{\mathbf E}_-^{(0)}$ as defined in subsection \ref{sec:AF}.
${\mathbf j}={\mathbf j}_-^{(1)}+{\mathbf j}_+^{(0)}+{\mathbf j}_-^{(0)}$
comprises all sources of the total magnetic field ${\mathbf B}$.
Note that the induced electric field does not contribute to the field energy in this approximation.

The field energy can be decomposed into three terms according to
\begin{eqnarray}\label{Bdecom}
&& \int \frac{1}{2\mu_0}\left|{\mathbf B}^{(1)}+{\mathbf B}^{(0)}\right|^2\,dV \\
\label{Bdecom1}
&=&  \int \frac{1}{2\mu_0}\left|{\mathbf B}^{(1)}\right|^2\,dV+
 \int \frac{1}{\mu_0}{\mathbf B}^{(1)}\cdot{\mathbf B}^{(0)}\,dV+
 \int \frac{1}{2\mu_0}\left|{\mathbf B}^{(0)}\right|^2\,dV\\
 \label{Bdecom2}
& =:& E_{field}^{11}+  E_{field}^{10}+ E_{field}^{00}
 \;.
 \end{eqnarray}
The first term $E_{field}^{11}$ is constant and can be neglected when considering energy balance.
For the time derivative of the second ``interference term" we obtain
\begin{equation*}
\frac{d}{dt}E_{field}^{10}=
{\textstyle \frac{3 \epsilon ^2 \dot{z}_0(t)}{2\left(\sqrt{2} \left(\mu _1+1\right)+24 \pi ^{3/2} \mu _2\right) }}
\end{equation*}
\begin{equation}\label{e10d}
{\textstyle \frac{ \left(\sqrt{2} \left(\mu _1+1\right) \omega _0
   \left(z_0(t)+1\right)^3-12 \pi ^{3/2} \mu _2 \left(-\mu _1+\left(\mu _1-2 \omega _0+1\right)
   z_0(t) \left(z_0(t) \left(z_0(t)+3\right)+3\right)-2 \omega _0-1\right)\right)}{\left(z_0(t)+1\right)^7}\;.}
\end{equation}
and, analogously,
\begin{eqnarray}\nonumber
&& \frac{d}{dt}E_{field}^{00}=
{\textstyle
-\frac{3 \left(\mu _1+1\right) \epsilon ^2 \dot{z}_0(t)}{\sqrt{2}
   \left(\sqrt{2} \left(\mu _1+1\right)+24 \pi ^{3/2} \mu _2\right)^2}
  } \\
   \label{e00d}
   &&
  {\textstyle \frac{\left(\sqrt{2} \left(\mu _1+1\right)
   \omega _0 \left(z_0(t)+1\right)^3+12 \pi ^{3/2} \mu _2 \left(2 \omega _0-\left(\mu _1-2
   \omega _0+1\right) z_0(t) \left(z_0(t) \left(z_0(t)+3\right)+3\right)\right)\right)}{\left(z_0(t)+1\right)^7}}
   \;.
  \end{eqnarray}

\begin{figure}[ht!]
\centering
\includegraphics*[clip,width=1.0\columnwidth]{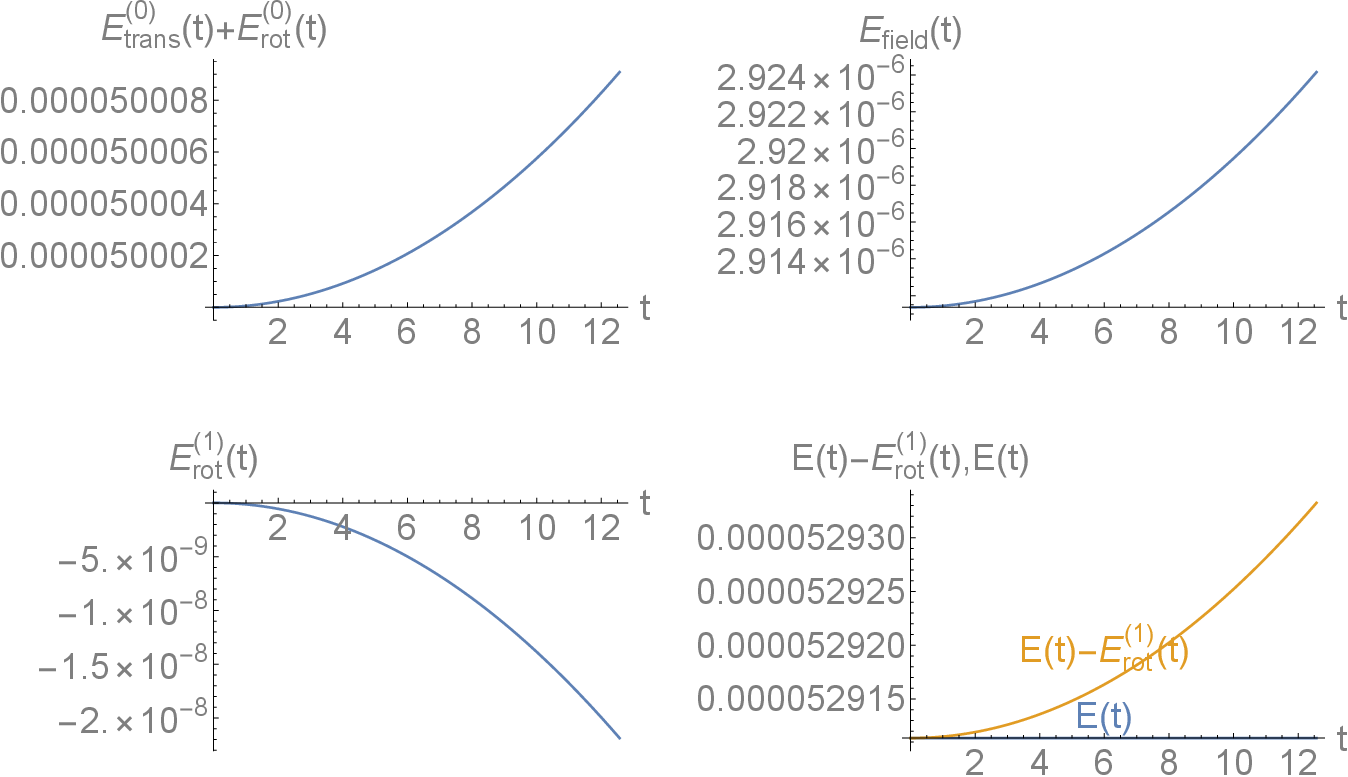}
\caption{Four quantities of the energy balance (\ref{energybalance}) are numerically calculated and shown as functions of $t$.
The total kinetic energy of the small dipole is slightly increasing (upper left panel) as well as  the field energy
$E_{field}:=E_{field}^{10}+E_{field}^{00}$
(upper right panel).
This is compensated by the decrease of rotational kinetic energy of the large dipole (lower left panel) such that the
total energy $E(t)$ is constant (lower right panel).
}
\label{FIGPE}
\end{figure}

Numerical examples show that $E_{field}^{00}$ decreases with time due to the weakening of the dipole moment
$m^{(0)}$ but this is overcompensated by the increase of $E_{field}^{10}$, see Figure \ref{FIGPE}.

For the two parts $E_{trans}^{(0)}$ and $E_{rot}^{(0)}$ of the kinetic energy of the small dipole we obtain
\begin{eqnarray}\nonumber
  &&\frac{d}{dt}E_{trans}^{(0)}=
  {\textstyle \frac{3 \epsilon ^2 \dot{z}_0(t)}{2 \left(z_0(t)+1\right)^7}}\\
  \label{etd}
  &&
  {\textstyle  \left(\frac{z_0(t) \left(z_0(t)
   \left(z_0(t)+3\right)+3\right) \left(\sqrt{2} \left(\mu _1+1\right) \omega
   _0-12 \pi ^{3/2} \mu_2 \left(\mu _1-2 \omega
   _0+1\right)\right)}{\sqrt{2} \left(\mu _1+1\right)+24 \pi ^{3/2} \mu
   _2}+\omega _0\right)}\;.
\end{eqnarray}
and
\begin{eqnarray}\nonumber
  &&\frac{d}{dt}E_{rot}^{(0)}=
  {\textstyle
  \frac{36 \pi ^{3/2} \mu _2 \epsilon ^2 \dot{z}_0(t)}{\left(\sqrt{2} \left(\mu _1+1\right)+24
   \pi ^{3/2} \mu _2\right)^2 }
   }
   \\
 \label{etr}
  &&
   {\textstyle
   \frac{
    \left(-\sqrt{2} \left(\mu
   _1+1\right) \omega _0 \left(z_0(t)+1\right){}^3-12 \pi ^{3/2} \mu _2 \left(2
   \omega _0-\left(\mu _1-2 \omega _0+1\right) z_0(t)
   \left(\left(z_0(t)+3\right)
   {z_0}(t)+3\right)\right)\right)}
   {\left(z_0(t)+1\right)^7}
   }
   \;.
   \end{eqnarray}

%===================    figure   =================================
%===================    figure  =======

Numerical examples and analytical considerations show that the sum of field energy and kinetic energy of the small dipole
is {\em not} constant, see Figure \ref{FIGPE}. Surprisingly, also the rotational kinetic energy $E_{rot}^{(1)}$ of the large dipole has to be taken into account.
We cannot calculate it directly, but its time derivative can be obtained by means of the torque ${\mathbf N}^{(01)}$
exerted by ${\mathbf E}^{(0)}$ upon the negatively charged cloud of the large dipole.
For this we use the equation $\frac{d}{dt}E_{rot}=\frac{d}{dt}{\small \frac{1}{2}}\theta\,\omega^2=\theta\,\omega\dot{\omega}=\omega\,N$
which also shows that the rotational energy of the positively charged cloud of the large dipole retains its initial value $0$
since $\omega_+^{(1)}=0$. Thus we obtain

\begin{eqnarray}\nonumber
  &&\frac{d}{dt}E_{rot}^{(1)}=
  {\textstyle
  \frac{3 \epsilon ^2 \dot{z}_0(t)}{2 \left(\sqrt{2}
   \left(\mu _1+1\right)+24 \pi ^{3/2} \mu _2\right)}
   }
   \\
\label{detr1}
  &&
    {\textstyle
    \frac{
     \left(12 \pi ^{3/2} \mu _2 \left(-\mu
   _1+\left(\mu _1-2 \omega _0+1\right) z_0(t) \left(z_0(t)
   \left(z_0(t)+3\right)+3\right)-2 \omega _0-1\right)-\sqrt{2} \left(\mu
   _1+1\right) \omega _0 \left(z_0(t)+1\right){}^3\right)}
   { \left(z_0(t)+1\right)^7}
   }
   \;.
 \end{eqnarray}

Finally, the following energy balance can be proven by computer-algebraic means:
\begin{prop}\label{P1}
 Under the preceding assumptions there holds
 \begin{equation}\label{energybalance}
  \frac{d}{dt}\left(E_{field}^{10}+E_{field}^{00}+E_{trans}^{(0)}+E_{rot}^{(0)} +E_{rot}^{(1)}\right)=0
  \;.
 \end{equation}
\end{prop}

We add a remark as to whether the change in rotational energy of the large dipole is compatible with the assumption of a constant external field.
Recall that $\frac{d}{dt}E_{rot}=\omega\,N$ and $\dot{\omega}=\frac{N}{\theta}$. If the inertial moment $\theta_-^{(1)}$ is very large compared with
$\theta_-^{(0)}$ and  $\theta_+^{(0)}$ then a finite value of $\frac{d}{dt}E_{rot}^{(1)}=\omega_-^{(1)}\,N^{(01)}$ will nevertheless lead to
a negligible value of $\dot{\omega}_-^{(1)}=\frac{N^{(01)}}{\theta_-^{(1)}}$. To this end it suffices to assume
\begin{equation}\label{largemassass}
  M_-^{(1)} \gg     M_-^{(0)},  M_+^{(0)}
  \;,
\end{equation}
thereby legitimizing talk of ``large" and ``small" dipole. This assumption also guarantees that the force exerted by the magnetic field of the small dipole on the large one (which has the same absolute value as the force exerted by the large dipole on the small one in accordance with the principle {\em actio} $=$ {\em reactio}) does not set the center of mass of the large dipole in motion. Both arguments together mean that the external field generated by the large dipole will be constant to an excellent approximation.

In order to illustrate the above results we have numerically solved the eom for the choice of parameters
$\epsilon=\mu_1=\mu_2=0.01$ and $\omega_0=\omega_1$. Some interesting terms of the energy balance (\ref{energybalance})
are shown as functions of $t$ in Figure \ref{FIGPE}.

%%%%%%%%%%%%%%%%%%%%%%%%%%%%%%%%%%%%%%%%%%%%%%%%%%%%%%%%%%%%%%%%%%%%%%%%%%%%%%%%%%%%%%%%%%%%%%%%%%%%%%%%%%%%%%%%%%%%%%%%%%%%%%%%%%%%%%%%%%%%%%%%%%
\section{Model B}
\label{sec:B}
%%%%%%%%%%%%%%%%%%%%%%%%%%%%%%%%%%%%%%%%%%%%%%%%%%%%%%%%%%%%%%%%%%%%%%%%%%%%%%%%%%%%%%%%%%%%%%%%%%%%%%%%%%%%%%%%%%%%%%%%%%%%%%%%%%%%%%%%%%%%%%%%%%

%%%%%%%%%%%%%%%%%%%%%%%%%%%%%%%%%%%%%%%%%%%%%%%%%%%%%%%%%%%%%%%%%%%%%%%%%%%%%%%%%%%%%%%%%%%%%%%%%%%%%%%%%%%%%%%%%%%%%%%%%%%%%%%%%%%%%%%%%%%%%%%%%%
\subsection{General notations and assumptions}
\label{sec:BG}
%%%%%%%%%%%%%%%%%%%%%%%%%%%%%%%%%%%%%%%%%%%%%%%%%%%%%%%%%%%%%%%%%%%%%%%%%%%%%%%%%%%%%%%%%%%%%%%%%%%%%%%%%%%%%%%%%%%%%%%%%%%%%%%%%%%%%%%%%%%%%%%%%%

In this section we model the small dipole by a rigid cloud carrying a genuine dipole density
\begin{equation}\label{dipoledensity}
 {\boldsymbol\mu}=
 {\mathbf m}^{(0)}\,\rho^{(0)} =
  \left(
\begin{array}{c}
 0 \\
 0 \\
 m^{(0)}  \\
\end{array}
\right)
\,\rho^{(0)}
\;,
\end{equation}
with $\rho^{(0)}=\rho(\left|{\mathbf r}-{\mathbf r}_0(t) \right|)$  given by (\ref{Gaussian}).
It turns out that the cloud will not rotate but will move along the $z$-axis. Hence the normalized density will
be written as $\rho^{(0)}=\rho(\left|(x,y,z-(a+z_0(t))) \right|)$, similarly as in Section \ref{sec:A}.
We largely adopt the notation of Section \ref{sec:A}.

Such genuine magnetic dipole densities have been considered in \cite{SB23} as additional sources of the electromagnetic
field resulting in the following extended form of Maxwell's equations:
\begin{eqnarray}
\label{modmax1}
\nabla\times {\bf E}+\frac{\partial}{\partial t}{\bf B}&=&{\bf 0},\\
\label{modmax2}
\nabla\cdot{\bf B}&=&0,\\
\label{modmax3}
\nabla\cdot\left({\epsilon_0\,\bf E}+{\mathbf Q}\right)&=&\rho,\\
\label{modmax4}
\frac{1}{\mu_0}\nabla\times {\mathbf B}-\frac{\partial}{\partial t}\left(\epsilon_0\,{\mathbf E}+{\mathbf Q}\right)
&=&
{\bf j}+\nabla\times {\boldsymbol{\mu}}
\;.
\end{eqnarray}
Here ${\mathbf Q}$ is a kind of ``quasi-electric polarization",
which must be added because of the relativistic covariance, but is not considered further in this paper.

Analogously to the usual electromagnetic theory a Poynting theorem can be derived from (\ref{modmax1} -- \ref{modmax4})
that describes the energy flow between field and matter:
\begin{equation}\label{PoyntingB}
 \frac{d}{d t} \int \left(\frac{1}{2\mu_0}\left| {\mathbf B}\right|^2+\frac{\varepsilon_0}{2}\left| {\mathbf E}\right|^2\right)\,dV =
  -\int\left( {\mathbf j}+\nabla\times{\boldsymbol\mu}+\frac{\partial}{\partial t}{\mathbf Q}\right)\cdot {\mathbf E}\,dV
 \;,
\end{equation}
where we have assumed that all fields decay sufficiently fast at infinity thereby neglecting radiation effects.

In general, a local magnetic moment ${\boldsymbol\mu} dV$ will precess around the external magnetic field ${\mathbf B}^{(1)}$.
This precession only does not occur on the $z$-axis, as the direction of ${\boldsymbol\mu}$ is parallel to ${\mathbf B}^{(1)}$.
If one deviates slightly from the $z$-axis, the direction of ${\boldsymbol\mu} dV$ is no longer parallel to ${\mathbf B}^{(1)}$
and the precession generates a position- and time-dependent direction field ${\mathbf m}^{(0)}({\mathbf r},t)$,
which replaces the constant direction field ${\mathbf m}^{(0)}$ in (\ref{dipoledensity}).
This effect, at least in linear order w.~r.~t.~${\mathbf r}$, should be taken into account for a complete description of the energy balance.
However, since it has no influence on the basic results, we move the ``inhomogeneity correction'' to  \ref{sec:A1}
and neglect this effect in the present section.

%%%%%%%%%%%%%%%%%%%%%%%%%%%%%%%%%%%%%%%%%%%%%%%%%%%%%%%%%%%%%%%%%%%%%%%%%%%%%%%%%%%%%%%%%%%%%%%%%%%%%%%%%%%%%%%%%%%%%%%%%%%%%%%%%%%%%%%%%%%%%%%%%%
\subsection{Force and torque}
\label{sec:BF}
%%%%%%%%%%%%%%%%%%%%%%%%%%%%%%%%%%%%%%%%%%%%%%%%%%%%%%%%%%%%%%%%%%%%%%%%%%%%%%%%%%%%%%%%%%%%%%%%%%%%%%%%%%%%%%%%%%%%%%%%%%%%%%%%%%%%%%%%%%%%%%%%%%

The large dipole is still realized by two massive rigid, charged clouds, so that only the negatively charged one rotates,
as in section \ref{sec:A}. As shown in \cite{SB23} the force density ${\mathbf f}$ within the small dipole due to the
external magnetic field ${\mathbf B}^{(1)}$ is given by
\begin{equation}\label{forcedensity}
 {\mathbf f} = \left(\nabla \times {\boldsymbol\mu}\right)\times {\mathbf B}^{(1)}
 \;,
\end{equation}
and the integration of (\ref{forcedensity}) over the whole space gives
\begin{eqnarray}\label{forceB1}
{\mathbf F}&=&\left.\left({\mathbf m}^{(0)}\cdot \nabla  \right){\mathbf B}^{(1)}\right|_{{\mathbf r}={\mathbf r}_0}\\
\label{forceB2}
&=&\left(0,0,-\frac{3\, B\,m^{(0)}}{a+z_0}\right)
  =\left(0,0,-\frac{3 m^{(0)} m^{(1)} \mu _0}{2 \pi  \left(a+z_0\right)^4}\right)
\;,
\end{eqnarray}
in accordance with (\ref{force}) and  eq.(37) of \cite{SB23}.

It can be shown that the total torque ${\mathbf N}^{(10)}$ exerted by  ${\mathbf B}^{(1)}$  on the small dipole vanishes, i.~e.,
\begin{equation}\label{zerotorque}
{\mathbf N}^{(10)}= \int {\mathbf r}\times \left( \left(\nabla \times {\boldsymbol\mu}\right)\times {\mathbf B}^{(1)}\right)\,dV ={\mathbf 0}
\;.
\end{equation}

%%%%%%%%%%%%%%%%%%%%%%%%%%%%%%%%%%%%%%%%%%%%%%%%%%%%%%%%%%%%%%%%%%%%%%%%%%%%%%%%%%%%%%%%%%%%%%%%%%%%%%%%%%%%%%%%%%%%%%%%%%%%%%%%%%%%%%%%%%%%%%%%%%
\subsection{Energy balance}
\label{sec:BB}
%%%%%%%%%%%%%%%%%%%%%%%%%%%%%%%%%%%%%%%%%%%%%%%%%%%%%%%%%%%%%%%%%%%%%%%%%%%%%%%%%%%%%%%%%%%%%%%%%%%%%%%%%%%%%%%%%%%%%%%%%%%%%%%%%%%%%%%%%%%%%%%%%%

The force (\ref{forceB1}) can be derived by ${\mathbf F}=-\nabla\,E_{Z}$ from a potential energy  $E_{Z}$, usually called ``Zeeman energy", given by
\begin{equation}\label{Zeeman}
  E_{Z}({\mathbf r}_0)=-{\mathbf m}^{(0)}\cdot  {\mathbf B}^{(1)}({\mathbf r}_0)=-\frac{m^{(0)}m^{(1)} \mu _0}{2 \pi  \left(a+z_0\right)^3}
  \;.
\end{equation}
To some extent it plays the role of the rotational kinetic energy of the small dipole in case A,
but now there holds a strict energy conservation law of the form
\begin{equation}\label{energyconservation0}
  \frac{d}{dt}\left(E_{trans}^{(0)}+E_Z \right)=0
  \;,
\end{equation}
since ${\mathbf m}^{(0)}$ is constant. Interestingly, the resulting equation of motion for the vertical coordinate $z_0(t)$
can be analytically solved, see \ref{sec:A2}.

To complete the energy balance we have to consider the field energy. Of the three terms of (\ref{Bdecom}--\ref{Bdecom2}) only the
``interference term" $E_{field}^{10}$ is not constant in time. From the comparison with model A we expect that the increase
of  $E_{field}^{10}$ is balanced by a corresponding decrease of $E_{rot}^{(1)}$. In fact, we can prove the following:
\begin{prop}\label{P2}
 Under the preceding conditions there holds:
 \begin{equation}\label{energyconservation1}
 \frac{d}{dt}\left(E_{field}^{10}+ E_{rot}^{(1)}\right)=0
 \end{equation}
\end{prop}

{\bf Proof}: We only consider the case without inhomogeneity corrections and
refer the reader to \ref{sec:A1} for the remaining part of the proof.
The magnetic field $ {\mathbf B}^{(0)}$ satisfies the quasi-static equations
$\nabla\cdot {\mathbf B}^{(0)} =0$ and $\nabla\times {\mathbf B}^{(0)}=\mu_0 \nabla\times{\boldsymbol\mu}$,
resulting from (\ref{modmax2}) and (\ref{modmax4}). It follows that
\begin{equation}\label{B0}
 {\mathbf B}^{(0)}= \mu_0\left({\boldsymbol\mu}+\nabla\, u\right)
 \;,
\end{equation}
with a scalar field $u$ satisfying
\begin{equation}\label{u}
 \Delta u = - \nabla\cdot {\boldsymbol\mu}
 \;,
\end{equation}
and vanishing at infinity. Then
\begin{eqnarray}
\label{E10a}
  E_{field}^{10} &=& \frac{1}{\mu_0} \int  {\mathbf B}^{(1)}\cdot  {\mathbf B}^{(0)} dV\\
  \label{E10b}
   &\stackrel{(\ref{B0})}{=}& \int  {\mathbf B}^{(1)}\cdot \left( {\boldsymbol\mu}+\nabla\,u  \right) dV\\
   \label{E10c}
   &=& \int  {\mathbf B}^{(1)}\cdot  {\boldsymbol\mu}\, dV+
   \int  {\mathbf B}^{(1)}\cdot \left( \nabla\,u  \right) dV
    \\
    \label{E10d}
   &=& \int  {\mathbf B}^{(1)}\cdot  {\boldsymbol\mu}\, dV+
   \int \left( \nabla\cdot\left({\mathbf B}^{(1)}u\right)-u\left(\nabla\cdot  {\mathbf B}^{(1)} \right)\right) dV\\
   \label{E10e}
   &=&\int  {\mathbf B}^{(1)}\cdot  {\boldsymbol\mu}\, dV\\
    \label{E10f}
   &\stackrel{(\ref{dipoledensity})}{=}& {\mathbf m}^{(0)}\cdot \int  {\mathbf B}^{(1)}\,\rho^{(0)}\, dV\\
    \label{E10g}
   &=& {\mathbf m}^{(0)}\cdot {\mathbf B}^{(1)}\left.\right|_{{\mathbf r}={\mathbf r}_0}
   \;,
\end{eqnarray}
using $\nabla\cdot {\mathbf B}^{(1)} =0$ and Gauss' integral theorem in (\ref{E10e})
as well as the point dipole approximation in (\ref{E10g}).
Interestingly, $E_{field}^{10}$ equals the {\em negative} Zeeman energy (\ref{Zeeman}) which shows that
$E_{field}^{10}$ also {\em increases} if the small dipole is accelerated and its kinetic energy increases,
similarly as in the model A.

In order to evaluate
\begin{equation}\label{dErot}
  \frac{d}{dt}E_{rot}^{(1)}= {\boldsymbol\omega}^{(1)}\cdot {\mathbf N}^{(01)}
\end{equation}
we consider
\begin{eqnarray}
\label{N1}
   {\mathbf N}^{(01)} &=& \int {\mathbf r}\times \left(e^{(1)}_-\rho^{(1)} {\mathbf E}^{(0)} \right)\,dV \\
   \label{N2}
  &\stackrel{(\ref{gradrho})}{=}&-\frac{e^{(1)}_-}{2\alpha^2}\int\nabla \rho^{(1)}\times  {\mathbf E}^{(0)} \,dV \\
  \label{N3}
  &=&-\frac{e^{(1)}_-}{2\alpha^2}\int\left(\nabla\times\left(\rho^{(1)}\, {\mathbf E}^{(0)} \right)-
  \left(\nabla\times {\mathbf E}^{(0)}\right)\,\rho^{(1)} \right) \,dV \\
  \label{N4}
    &=&\frac{e^{(1)}_-}{2\alpha^2}\int \left(\nabla\times {\mathbf E}^{(0)}\right)\,\rho^{(1)} \,dV \\
    \label{N5}
    &\stackrel{(\ref{modmax1})}{=}& -\frac{e^{(1)}_-}{2\alpha^2}\int\left(\frac{d}{dt} {\mathbf B}^{(0)}\right)\,\rho^{(1)}\,dV\\
    \label{N6}
    &=&  -\frac{e^{(1)}_-}{2\alpha^2}\left.\left(\frac{d}{dt} {\mathbf B}^{(0)}\right)\right|_{{\mathbf r}={\mathbf 0}}
       \;,
\end{eqnarray}
where we have used Gauss' integral theorem in (\ref{N4}) and the point dipole approximation in (\ref{N6}).
Hence
\begin{equation}\label{dErota}
 \frac{d}{dt}E_{rot}^{(1)}\stackrel{(\ref{dErot},\ref{N6})}{=}
  \frac{d}{dt} \left(\left. - {\mathbf m}^{(1)}\cdot{\mathbf B}^{(0)}\right|_{{\mathbf r}={\mathbf 0}}\right)
 \;.
\end{equation}
The claim (\ref{energyconservation1}) then follows from
\begin{equation}\label{Bid}
 \left. {\mathbf m}^{(1)}\cdot{\mathbf B}^{(0)}\right|_{{\mathbf r}={\mathbf 0}} =
 {\mathbf m}^{(0)}\cdot {\mathbf B}^{(1)}\left.\right|_{{\mathbf r}={\mathbf r}_0} =m^{(0)} m^{(1)} \frac{\mu_0}{4\pi}\frac{2}{\left( a+z_0\right)^3}
 \;.
\end{equation}
\hfill$\Box$\\

%%%%%%%%%%%%%%%%%%%%%%%%%%%%%%%%%%%%%%%%%%%%%%%%%%%%%%%%%%%%%%%%%%%%%%%%%%%%%%%%%%%%%%%%%%%%%%%%%%%%%%%%%%%%%%%%%%%%%%%%%%%%%%%%%%%%%%%%%%
\section{Summary}\label{sec:S}
%%%%%%%%%%%%%%%%%%%%%%%%%%%%%%%%%%%%%%%%%%%%%%%%%%%%%%%%%%%%%%%%%%%%%%%%%%%%%%%%%%%%%%%%%%%%%%%%%%%%%%%%%%%%%%%%%%%%%%%%%%%%%%%%%%%%%%%%%%

The problem of the interaction between two magnetic dipoles is interesting in itself
and also as a case study to investigate the title question of whether the magnetic field can perform work.
In this paper we have considered the special case in which both magnetic moments are
initially aligned parallel to the line connecting the positions of the two dipoles,
and, moreover, the limiting case in which the ``large" dipole is so massive
that it practically does not move and generates an external magnetic field, constant in time,
in which the ``small" dipole is accelerated.

For the detailed investigation, we have considered two cases, which we refer to as cases A and B.
In the realm of Maxwell theory, it may be a matter of taste
whether the magnetic moments are simulated by currents (model A) or
considered as genuine quantities (model B). But for the energy balance
we have to consider mechanical models for the dipoles,
and then it makes a considerable difference which model is used.

We have derived energy balance equations for both models.
For case A, the sum of the total kinetic energy of the two dipoles and the field energy is constant.
Surprisingly, the rotational kinetic energy of the large dipole must also be taken into account
as the source of the increase in both the field energy and the total kinetic energy of the small dipole.
For case B, the role of the rotational kinetic energy of the small dipole is largely taken over by the Zeeman energy.
In contrast to model A, here we have two separate energy conservation equations,
one for the small dipole and another one for the field energy and the large dipole.

However, the question whether the magnetic field can do work
cannot be answered unambiguously, because it is not clear what it
means exactly.
For example, J.~A.~Barandes argues that the magnetic field can do
work since in his approach an energy balance analogous to our
(\ref{energyconservation0})
is satisfied, see \cite{B21}, eq.~(40). On the other hand, the r.~h.~s.~of the
Poynting theorem (\ref{PoyntingB}), which summarizes the energy flow from matter
to the electromagnetic field, is linear in ${\mathbf E}$ and does not contain ${\mathbf B}$, also in the
extended version based on \cite{SB23}. Hence the magnetic field
energy cannot be transformed into matter energy unless ${\mathbf E}\neq{\mathbf 0}$.

It is instructive to compare the repulsion of two magnetic dipoles with the repulsion
of two charges of the same sign, which are similarly modeled by rigid charge clouds of extremely different mass.
Here it is certainly legitimate to say that the electric field of the "large" charge performs work, according to whatever criteria.
The increase in kinetic energy of the ``small" charge is compensated for by a decrease in potential energy, analogous to our case B.
However, in the case of electric charges, the potential energy can be interpreted as the field energy of the total electric field.
This field energy therefore {\em decreases}, while the magnetic field energy {\em increases} in both of our cases, A and B.
In our opinion, the detailed description of the differences between the two examples of repulsion is much more revealing
than the reduction to the simple alternative ``the magnetic field performs work - or it does not".

%%%%%%%%%%%%%%%%%%%%%%%%%%%%%%%%%%%%%%%%%%%%%%%%%%%%%%%%%%%%%%%%%%%%%%%%%%%%%%%%%%%%%%%%%%%%%%%%%%%%%%%%%%%%%%%%%%%%%%%%%%%%%%%%%%%%%%%%%%
\section*{Acknowledgment}\label{sec:ACK}
%%%%%%%%%%%%%%%%%%%%%%%%%%%%%%%%%%%%%%%%%%%%%%%%%%%%%%%%%%%%%%%%%%%%%%%%%%%%%%%%%%%%%%%%%%%%%%%%%%%%%%%%%%%%%%%%%%%%%%%%%%%%%%%%%%%%%%%%%%
We thank Bruno Klajn for bringing the topic of this paper and recent literature to our attention,
and Jacob A. Barandes for insightful discussions and comments on earlier versions of the manuscript.
H.-J.~S.~would also like to express his gratitude to J\"urgen Schnack for organizing a
mini-symposium at the University of Bielefeld, where he was allowed to present parts of the present work.

\appendix

%%%%%%%%%%%%%%%%%%%%%%%%%%%%%%%%%%%%%%%%%%%%%%%%%%%%%%%%%%%%%%%%%%%%%%%%%%%%%%%%%%%%%%%%%%%%%%%%%%%%%%%%%%%%%%%%%%%%%%%%%%%%%%%%%%%%%%%%%%
\section{Inhomogeneity correction }\label{sec:A1}
%%%%%%%%%%%%%%%%%%%%%%%%%%%%%%%%%%%%%%%%%%%%%%%%%%%%%%%%%%%%%%%%%%%%%%%%%%%%%%%%%%%%%%%%%%%%%%%%%%%%%%%%%%%%%%%%%%%%%%%%%%%%%%%%%%%%%%%%%%

We replace (\ref{dipoledensity}) by
\begin{equation}\label{inhomdensity}
 {\boldsymbol\mu}=
{\mathbf m}_I \,\rho :=
  \left(
\begin{array}{c}
 I_1(t) x+I_2(t) y \\
 -I_2(t) x+I_1(t) y \\
 m^{(0)}  \\
\end{array}
\right)
\,\rho(|{\mathbf r}-{\mathbf r}_0(t)|)
\;,
\end{equation}
taking into account the $SO(2)$ symmetry w.~r.~t.~rotations around the $z$-axis and the
condition that the direction field ${\mathbf m}_I$ has the same absolute value
everywhere in the linear order $O({\mathbf r})$. The integral over the linear parts in
$ {\boldsymbol\mu}$ vanishes so that the total magnetic moment is unchanged:
\begin{equation}\label{intinhodens}
  \int  {\boldsymbol\mu}\,dV= {\mathbf m}^{(0)} = \left(
\begin{array}{c}
 0\\
0 \\
 m^{(0)}  \\
\end{array}
\right)
\;.
\end{equation}
Consequently, the equation for the force on the small dipole (\ref{forceB1}) still holds.
Also the total torque vanishes, as in (\ref{zerotorque}), due to  $SO(2)$ symmetry.
Hence the first energy balance equation (\ref{energyconservation0}) is left unchanged.

We introduce the ``gyromagnetic ratio" $\chi$ as the proportionality factor between moment density
${\boldsymbol\mu}$ and spin density ${\boldsymbol\sigma}$, i.~e.,
\begin{equation}\label{chi}
{\boldsymbol\mu} = \chi\, {\boldsymbol\sigma}
\;.
\end{equation}
Then the precession of ${\boldsymbol\mu}$ around the local magnetic field ${\mathbf B}^{(0)}$
is given by the equation
\begin{equation}\label{precession}
 \frac{d}{dt}{\boldsymbol\mu} = \chi\, {\boldsymbol\mu}\times {\mathbf B}^{(0)}
 \;.
\end{equation}
This entails the equations of motion for the functions $I_1(t)$ and $I_2(t)$ which read in first order $O({\mathbf r})$:
\begin{eqnarray}
\label{eomM1}
  \frac{d}{dt}I_1(t) &=& -\frac{m^{(1)} \mu _0 \chi  I_2(t)}{2 \pi \left(a+z_0(t)\right)^3}\;, \\
   \frac{d}{dt}I_2(t) &=& -\frac{3 m^{(0)} m^{(1)} \mu _0 \chi }{4 \pi  \left(a+z_0(t)\right)^4}
   +\frac{m^{(1)} \mu _0 \chi  I_1(t)}{2 \pi  \left(a+z_0(t)\right)^3}
   \;.
\end{eqnarray}

It remains to calculate the inhomogeneity corrections to $\frac{d}{dt}E_{field}^{10}$ and to
$\frac{d}{dt} E_{rot}^{(1)}$. For  $E_{field}^{10}$ we obtain additional terms in the integral (\ref{E10f})
of order $O\left({\mathbf r}^2\right)$ times $\rho^{(0)}$ which result in:
\begin{equation}\label{E10ic}
E_{field}^{10}={\mathbf m}^{(0)}\cdot {\mathbf B}^{(1)}({\mathbf r}_0)+\frac{3 m^{(1)} \mu_0 I_1(t)}{4 \pi  \alpha ^2 \left(a+z_0(t)\right)^4}
\;.
\end{equation}
The time derivative of  $E_{field}^{10}$ hence reads:
\begin{equation}\label{dE10ic}
\frac{d}{dt}E_{field}^{10}=
{\textstyle
{\mathbf m}^{(0)}\cdot  \frac{d}{dt}{\mathbf B}^{(1)}({\mathbf r}_0(t))
+\frac{3 m^{(1)} \mu _0 \left(\left(a+z_0(t)\right) \dot{I}_1(t)-
4 I_1(t) \dot{z}_0(t)\right)}{4 \pi\alpha ^2 \left(a+z_0(t)\right)^5}
}
\;.
\end{equation}

For  $\frac{d}{dt} E_{rot}^{(1)}$ we have to calculate the inhomogeneity correction to the magnetic field ${\mathbf B}^{(0)}$.
To this end we recall (\ref{B0}) and (\ref{u}) and note that
\begin{equation}\label{divmu}
  \nabla\cdot {\boldsymbol\mu}=-2 \rho^{(0)}  \left(I_1 \left(-1+\alpha ^2 \left(x^2+y^2\right)\right)+m^{(0)} \alpha ^2 u\right)
  \;.
\end{equation}
We consider the well-known equation
\begin{equation}\label{Laplacerho}
\Delta\left(-\frac{1}{4\pi r}\mbox{erf}(\alpha r) \right)=\rho
\end{equation}
and apply differential operators $\frac{\partial}{\partial x}, \frac{\partial}{\partial y}, \ldots$ on both sides,
using that these differential operators commute with $\Delta$. This gives polynomials of $x,y,z$ times $\rho$
at the r.~h.~s., and thus we can obtain a solution of $ \Delta u = - \nabla\cdot {\boldsymbol\mu}$, and,
further the magnetic field ${\mathbf B}^{(0)}$ with inhomogeneity correction. However, it is too complex
to be shown here. Following the arguments that lead from (\ref{N1}) to (\ref{N6}) we finally obtain
\begin{equation}\label{Erotic}
 \frac{d}{dt}E_{rot}^{(1)}{=}
  {\textstyle
  \frac{d}{dt} \left(\left. -
 {\mathbf m}^{(1)}\cdot{\mathbf B}^{(0)}\right|_{{\mathbf r}={\mathbf 0}}\right)
 -\frac{3 m^{(1)} \mu _0 \left(\left(a+z_0(t)\right) \dot{I}_1(t)-4 I_1(t) \dot{z}_0(t)\right)}
 {4 \pi  \alpha ^2 \left(a+z_0(t)\right)^5}
 }
 \;.
\end{equation}
Comparison of (\ref{Erotic}) and (\ref{dE10ic}) proves that the energy balance
(\ref{energyconservation1}) also holds if taking the inhomogeneity correction into account.

%%%%%%%%%%%%%%%%%%%%%%%%%%%%%%%%%%%%%%%%%%%%%%%%%%%%%%%%%%%%%%%%%%%%%%%%%%%%%%%%%%%%%%%%%%%%%%%%%%%%%%%%%%%%%%%%%%%%%%%%%%%%%%%%%%%%%%%%%%
\section{Analytical solution of the eom for model B }\label{sec:A2}
%%%%%%%%%%%%%%%%%%%%%%%%%%%%%%%%%%%%%%%%%%%%%%%%%%%%%%%%%%%%%%%%%%%%%%%%%%%%%%%%%%%%%%%%%%%%%%%%%%%%%%%%%%%%%%%%%%%%%%%%%%%%%%%%%%%%%%%%%%

%===================    figure   =================================
\begin{figure}[ht!]
\centering
\includegraphics*[clip,width=0.7\columnwidth]{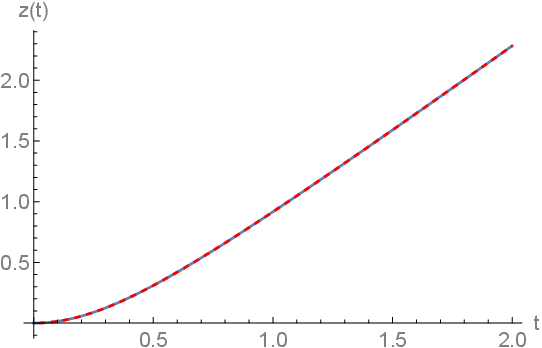}
\caption{Numerical solution of the eom (\ref{eomB}) with the special value of $\lambda=1$  (blue curve)
and analytical solution according to (\ref{t4}) (red, dashed curve).
}
\label{FIG12}\end{figure}
%===================    figure  =======

Eq.~(\ref{precession}) shows that $\omega_1= \chi B$ is a natural unit for the angular velocity in case of model B. It can be
identified with $\omega_1$ defined by (\ref{Larmor}) for model A if we introduce a formal charge
\begin{equation}\label{defe}
  e^{(0)}:=\chi\,M^{(0)}
\end{equation}
also for model B and identify $M^{(0)}$ with $M^{(0)}_-$.
Then we can adopt the dimensionless quantities from Section \ref{sec:AE}
and write the eom for the model B in the form
\begin{equation}\label{eomB}
 \ddot{z}_0(t)= 3 \underbrace{\left(2 \pi |\gamma_1| \frac{\mu_2}{\epsilon}\right)}_{=:\lambda}\,\frac{1}{(1+z_0(t)^4)}
 \;.
\end{equation}
Using these quantities the first energy balance (\ref{energyconservation0}) assumes the form
\begin{equation}\label{econ1}
 \frac{1}{2}\dot{z}_0(t)^2 +\frac{\lambda}{2}\frac{1}{(1+z_0(t)^3)} =E
 \;.
\end{equation}
With the initial conditions $z_0(0)=0$ and $\dot{z}_0(0)=0$ we obtain for the total energy
$E=\frac{\lambda}{2}$ and hence
\begin{equation}
\label{econ2}
  \dot{z}_0(t)^2 = \lambda\left(1-\frac{1}{(1+z_0(t))^3} \right)
  \;.
  \label{econ2b}
 \end{equation}
Obviously, for $t\to \infty$ we have $z_0(t)\to \infty$ and $\dot{z}_0(t)\to \sqrt{\lambda}$.
Solving (\ref{econ2}) for $t$ yields
 \begin{eqnarray} \label{t1}
  t&=&\int_{0}^{z_0}\frac{dz}{\sqrt{\lambda\left(1-\frac{1}{(1+z)^3}\right)}}\\
   \label{t2}
  &=& \frac{-1}{3\sqrt{\lambda}}\int_{1}^{(1+z_0)^{-3}} u^{-4/3} (1-u)^{-1/2}\,du
    \\
    \label{t3}
  &=&   \frac{-1}{3\sqrt{\lambda}} \left( B\left(-{\textstyle \frac{1}{3}},{\textstyle \frac{1}{2}},(1+z_0)^{-3}\right)
  -B\left(-{\textstyle \frac{1}{3}},{\textstyle \frac{1}{2}},1\right)\right)\\
    \label{t4}
  &=& \frac{1}{\sqrt{\lambda}}(1+z_0) \left( _2F_1\left(-{\textstyle \frac{1}{3}},{\textstyle \frac{1}{2}};{\textstyle \frac{2}{3}};(1+z_0)^{-3}\right)-
 \frac{\sqrt{\pi } \Gamma \left(\frac{2}{3}\right)}{\Gamma \left(\frac{1}{6}\right)}\right)
  \;.
\end{eqnarray}
In (\ref{t2}) we have substituted $u=(1+z)^{-3}$ in order to write the integral in terms of the incomplete beta function
$B(a,b,x)$, see \cite{NIST}, (8.17.1). The latter can also be expressed
by the hypergeometric function $_2F_1\left(a,b;c; x\right)$., see \cite{NIST}, (8.17.7), which gives (\ref{t4}).
Thus we have obtained an analytical solution of the eom in
terms of the inverse function $t\mapsto t(z_0)$.

We have compared this solution with a numerical solution of the eom choosing the special value of $\lambda=1$
and found perfect agreement, see Figure \ref{FIG12}.

%%%%%%%%%%%%%%%%%%%%%%%%%%%%%%%%%%%%%%%%%%%%%%%%%%%%%%%%%%%%%%%%%%%%%%%%%%%%%%%%%%%%%%%%%%%%%%%%%%%%%%%%%%%%%%%%%%%%%%%%%%%%%%%%%%%%%%%%%%%%%%%%%%
\section*{References}
%%%%%%%%%%%%%%%%%%%%%%%%%%%%%%%%%%%%%%%%%%%%%%%%%%%%%%%%%%%%%%%%%%%%%%%%%%%%%%%%%%%%%%%%%%%%%%%%%%%%%%%%%%%%%%%%%%%%%%%%%%%%%%%%%%%%%%%%%%%%%%%%%%

%%%%%%%%%%%%%%%%%%%%%%%%%%%%%%%%%%%%%%%%%%%%%%%%%%%%%%%%%%%%%%%%%%%%%%%%%%%%%%%%%%%%%%%%%%%%%%%%%%%%%%%%%%%%%%%%%%%%%%%%%%%%%%%%%%%%%%%%%%%%%%%%%%

\begin{thebibliography}{15}

\bibitem{J90}
J.~D.~Jackson,
\textit{Classical Electrodynamics}, 3rd edition, Wiley, New York, 1999.


\bibitem{M74}
E.~P.~Mosca,
Magnetic Forces Doing Work?,
{\textit Am. J. Phys.} {\bf 42}, 295 --297 (1974)


\bibitem{C79}
C.~A.~Coombes,
Work done on charged particles in magnetic fields,
{\textit Am. J. Phys.} {\bf 47}, 915 -- 916 (1979)


\bibitem{OA13}
P.~Onorato and A.~De Ambrosis,
How can magnetic forces do work? Investigating the problem with students,
{\textit Phys. Educ.} {\bf 48} (6), 766 (2013)

\bibitem{G14}
J.~Gates,
Magnetic force and work: an accessible example,
{\textit Phys. Educ.} {\bf 49} (3), 299 -- 302 (2014)


\bibitem{VSR23}
S.~C.~Veetil, H.~X.~Sim, and B.~Ricardo,
Magnetic Forces: Do They Really Work?,
{\textit The Phys. Educ.} {\bf 5} (1), 2350001 (2023)

\bibitem{B19}
J.~A.~Barandes,
Can magnetic forces do work?,
\textit{Preprint} arXiv:1911.08890 (2019)


\bibitem{B21}
J.~A.~Barandes,
On magnetic forces and work,
{\textit Found. Phys.} {\bf 51} (4), 79, (2021)


\bibitem{SB23}
H.-J.~Schmidt and T.~Br\"ocker,
 Maxwell equations with spin density,
{\textit Eur. J. Phys.} {\bf 44}(3), 035201, (2023)


\bibitem{SSHL15}
H.-J.~Schmidt, C.~Schr\"oder, E.~H\"agele, and M.~Luban,
Dynamics and thermodynamics of a pair of interacting magnetic dipoles,
{\textit J. Phys. A, Math. Theor.} {\bf 48}(18), 185002, (2015)


\bibitem{BD64}
J.~D.~Bjorken and S.~D.~Drell,
\textit{Relativistic Quantum Mechanics}, McGraw-Hill, New York, 1964.


\bibitem{M82}
J.~S.~Marsh,
Magnetic and electric fields of rotating charge distributions,
{\textit Am. J. Phys.} {\bf 50}, 51 -- 53, (1982)


\bibitem{B88}
T.~H.~Boyer,
The force on a magnetic dipole,
{\textit Am. J. Phys.} {\bf 56}, 688 -- 692, (1988)

\bibitem{V90}
L.~Vaidman,
Torque and force on a magnetic dipole,
{\textit Am. J. Phys.} {\bf 58}, 978 -- 983, (1990)


\bibitem{NIST}
NIST Digital Library of Mathematical Functions. http://dlmf.nist.gov/,
Release 1.1.1 of 2023-09-15.
F.~W.~J.~Olver, A.~B.~Olde Daalhuis, D.~W.~Lozier, B.~I.~Schneider, R.~F.~Boisvert,
C.~W.~Clark, B.~R.~Miller, B.~V.~Saunders, H.~S.~Cohl, and M.~A.~McClain, eds.

\end{thebibliography}
\end{document}